\pdfoutput=1

\documentclass[11pt,twoside,a4paper,cmspaper,final,collab]{cms-tdr}
\def\svnVersion{231014}\def\cmsCernNo{2013-037}\def\cmsCernDate{\today}\def\cmsMessage{Published in Physics Letters B as \href{http://dx.doi.org/10.1016/j.physletb.2013.02.031}{\doi{10.1016/j.physletb.2013.02.031.}}}

\begin{document}\cmsNoteHeader{EXO-11-034}

\hyphenation{had-ron-i-za-tion}
\hyphenation{cal-or-i-me-ter}
\hyphenation{de-vices}
\RCS$Revision: 171142 $
\RCS$HeadURL: svn+ssh://alverson@svn.cern.ch/reps/tdr2/papers/EXO-11-034/trunk/EXO-11-034.tex $
\RCS$Id: EXO-11-034.tex 171142 2013-02-12 12:09:29Z alverson $

\newlength\cmsFigWidth
\ifthenelse{\boolean{cms@external}}{\setlength\cmsFigWidth{0.9\columnwidth}}{\setlength\cmsFigWidth{0.6\textwidth}}
\ifthenelse{\boolean{cms@external}}{\providecommand{\cmsLeft}{top}}{\providecommand{\cmsLeft}{left}}
\ifthenelse{\boolean{cms@external}}{\providecommand{\cmsRight}{bottom}}{\providecommand{\cmsRight}{right}}
\providecommand\mustar{\ensuremath{\mu^*}\xspace }
\providecommand\estar{\ensuremath{\Pe^*}\xspace }
\providecommand\lstar{\ensuremath{\ell^*}\xspace }
\providecommand\llg{\ensuremath{\ell^+ \ell^- \gamma}\, }
\providecommand\Mmin{\ensuremath{M_{\ell \gamma}^\text{min}}\xspace }
\providecommand\Mmax{\ensuremath{M_{\ell \gamma}^\text{max}}\xspace }
\def\etiso{\ensuremath{\mathit{E_\mathrm{T}^\text{iso}}}\xspace}
\def\phantomdagger{\,\phantom{$^{\dagger}$}}

\providecommand\newrb{\relax}\providecommand\newre{\relax}\providecommand\newb{\relax}\providecommand\newe{\relax}\providecommand\newbb{\relax}\providecommand\newbe{\relax}
\cmsNoteHeader{EXO-11-034} % This is over-written in the CMS environment: useful as preprint no. for export versions
\title{Search for excited leptons in pp collisions at $\sqrt{s} = 7\TeV$}
\date{\today}
\abstract{
Results are presented of a search for compositeness in electrons and muons using a  data sample of pp
collisions at a
center-of-mass energy $\sqrt{s}$ = 7\TeV collected with the CMS detector
at the LHC and corresponding to an integrated luminosity of 5.0\fbinv.
Excited leptons ($\ell^*$) are
assumed to be produced via contact interactions
in conjunction with a standard model lepton and to decay via
$\ell^* \to \ell \gamma$, yielding a final state with two energetic leptons and a photon.
The number of events observed in data is consistent with that expected from the standard model.
The 95\% confidence upper limits for the cross section
for the production and decay of excited electrons (muons),
with masses ranging from 0.6 to 2\TeV, are
1.48 to 1.24\unit{fb} (1.31 to 1.11\unit{fb}).
Excited leptons with masses below 1.9\TeV are excluded for the
case where the contact interaction scale
equals the excited lepton mass.
The limits on the cross sections are the most stringent ones published to date.
}
\hypersetup{%
pdfauthor={CMS Collaboration},%
pdftitle={Search for excited leptons in pp collisions at sqrt(s) = 7 TeV},%
pdfsubject={CMS},%
pdfkeywords={CMS, physics}}
\maketitle %maketitle comes after all the front information has been supplied

\section{Introduction}
\label{sec:intro}

The standard model (SM) of particle physics, albeit very successful,
provides no explanation for the three generation
structure of the fermion families.
Attempts to explain the observed hierarchy have
led to a family of
models postulating that  quarks and leptons might be composite objects of
fundamental constituents
\cite{Pati:1975md,compositeness1, Eichten1982, Eichten:1983hw, harari, ssc-physics, Baur90,
Greenberg:1974qb, Greenberg:1980ri}.
The fundamental constituents are bound by an asymptotically free
gauge interaction
that becomes strong at a characteristic scale $\Lambda$.
Compositeness models predict
the existence of excited states of quarks (q$^{*}$) and leptons (\lstar)
at this characteristic scale of
the new binding interaction.
Since these excited fermions couple to the ordinary SM
fermions, they can be produced
via contact interactions in collider experiments and
subsequently decay
radiatively to ordinary fermions through the emission of a W/Z/$\gamma$ boson
or  via contact  interactions to other fermions.
The excited leptons can also be produced via  gauge-mediated interactions, but the
cross sections for these are negligible for the range of
parameters that are probed in this search and therefore
this production mechanism is not considered.
The effective Lagrangian describing the interaction of excited fermions~\cite{Baur90} is
parametrized by the scale  $\Lambda$. Additionally, \newbb for decay via gauge mediated interaction, \newbe
two factors $f$ and $f'$ represent the relative strength of the coupling between the excited
fermions and isovector and isoscalar gauge fields, respectively. In this Letter the convention  $f=f'=1$ is adopted. The
results for arbitrary $f=f'>0$ can be simply obtained by a rescaling of the scale $\Lambda$ to $\Lambda/f$.

Searches    at   LEP~\cite{Buskulic:1996tw,   Abreu:1998jw,
Abbiendi:1999sa, Achard:2003hd},    HERA~\cite{H1estar},   and   the
Tevatron~\cite{CDFestar,cdfmu,d0,D0estar} found no evidence for
excited  leptons.
At  the Large  Hadron
Collider (LHC)~\cite{LHC} at CERN,
previous searches  performed  by the
CMS~\cite{cms-limit} and the ATLAS collaborations~\cite{atlas-limit}
have also shown  no evidence  for excited  leptons.
At  a  center-of-mass energy of $\sqrt{s}$ = 7\TeV, with 36\pbinv of data ~\cite{cms-limit}, CMS has excluded
cross sections for the production and decay of  the $\ell^* \rightarrow \ell \gamma$ channels higher than
0.16 to 0.21\unit{pb} (0.14 to 0.19\unit{pb})
in the \estar (\mustar) channel for excited lepton masses ranging from 0.2\TeV to 2\TeV. In the same
channels and with more integrated luminosity, ATLAS excluded cross sections higher than
2.3~(4.5)\unit{fb}
for excited electrons (muons) masses
above 0.9\TeV, and excluded
\estar (\mustar) with masses $M_{\ell^*}$ below 1.87~(1.75)\TeV for the scale of contact interaction $\Lambda =
M_{\ell^*}$~\cite{atlas-limit}.

This Letter presents a search for excited leptons, \estar and \mustar,
using a data sample of pp collisions at a center-of-mass energy $\sqrt{s}$ = 7\TeV collected with the CMS
detector at the LHC
 in 2011
and corresponding to an integrated luminosity of $5.0\pm0.1\fbinv$.
The production of an excited lepton in association
with an oppositely charged lepton of the same flavor,
via four-fermion contact interactions,
is considered.
 Thus when the excited lepton decays via $\ell^* \rightarrow \ell \gamma$,
there are two oppositely charged leptons and a photon in the final state.

\section{The CMS detector}
\label{sec:detector}

The central feature of the Compact Muon Solenoid (CMS)
detector is a superconducting solenoid, of
6\unit{m} internal diameter and 12.5\unit{m} in length, which provides
an axial field of 3.8\unit{T}.
Starting from the collision point, the first three
detector components inside the solenoid are the silicon
pixel and strip trackers; the lead-tungstate crystal electromagnetic
calorimeter (ECAL), comprising a central (barrel) section and two  forward (endcap) sections; and the
brass/scintillator hadron calorimeter  (HCAL). Extensive forward calorimetry complements the coverage provided
by the barrel and endcap detectors.
The tracker consists of 10 layers of silicon strip detectors in addition to the pixel detectors.
Four stations of muon detectors are embedded in the steel yoke of the superconducting solenoid,
 including forward sections in order to extend the covered pseudorapidity region up
to $|\eta| < 2.4$. The pseudorapidity ($\eta$) is defined as $\eta = -\ln[\tan(\theta/2)]$.
The CMS detector uses a right-handed coordinate system,
with  the origin at the nominal  interaction point, the $x$ axis pointing to  the
center of the LHC,
the  $y$ axis pointing up (perpendicular to the  LHC plane), and the $z$ axis
along the anticlockwise-beam direction. The
polar angle $\theta$ is
measured from the positive $z$ axis and the azimuthal angle $\phi$ is
measured in the $x$-$y$ plane.
The projection of the momentum on to the $x$-$y$ plane is used to define the
transverse momentum \pt and the transverse energy \et.
The details of the CMS detector are described elsewhere~\cite{JINST}.

\section{Signal and background}
\label{sec:samples}

The dominant, irreducible SM background in this search is Drell--Yan production of
$\ell^+\ell^-\gamma$ where the final state
photon is either radiated by an initial-state parton (initial-state radiation, ISR), or originates from one
of the final-state leptons (final-state radiation, FSR). The second-most important background is due to
Drell--Yan production associated with jets (Z+jets), where a jet is misidentified as a photon
(see Section~\ref{sec:bkg}).
Another important background in the $\estar$ channel is due to W+jets
events with an FSR or ISR photon where a jet is misidentified as an electron.
In the $\mustar$ channel, backgrounds from these W+jets processes that lead
to one true, one misidentified
 muon, and a true photon in the final state have been estimated to be negligible.
Other less significant backgrounds originate from diboson events (WW, WZ, ZZ, \newbb $\PW + \gamma$ \newbe),
$\ttbar$ production, and, for the electron channel, $\gamma\gamma$ production. These backgrounds are
mainly suppressed by
requiring high transverse momentum
thresholds on the leptons
and photon.
Backgrounds arising from misidentified photons or misidentified electrons
are estimated using a data-driven technique which is described in
Section \ref{sec:bkg}. The other backgrounds are estimated from the simulation.

Signal samples in both electron and muon channels are
 produced using {\PYTHIA} ({\PYTHIA
6.424}~\cite{Sjostrand:2006za} and {\PYTHIA
8.145}~\cite{Sjostrand:pythia8} respectively)
 based on the leading order (LO)
compositeness model described in Ref.~\cite{Baur90}.
 The signal cross sections are calculated with {\PYTHIA 6.424}, corrected to include
the branching ratio for the 3-body decays \newbb via contact interaction as per Ref.~\cite{Baur90} \newbe
which is not implemented in {\PYTHIA}, with the $Q^{2}$ scale set to the square of the mass of the excited
lepton ($M_{\ell^*}^{2}$).

\begin{figure}[th!bp]
\begin{center}
\includegraphics[width=0.45\textwidth]{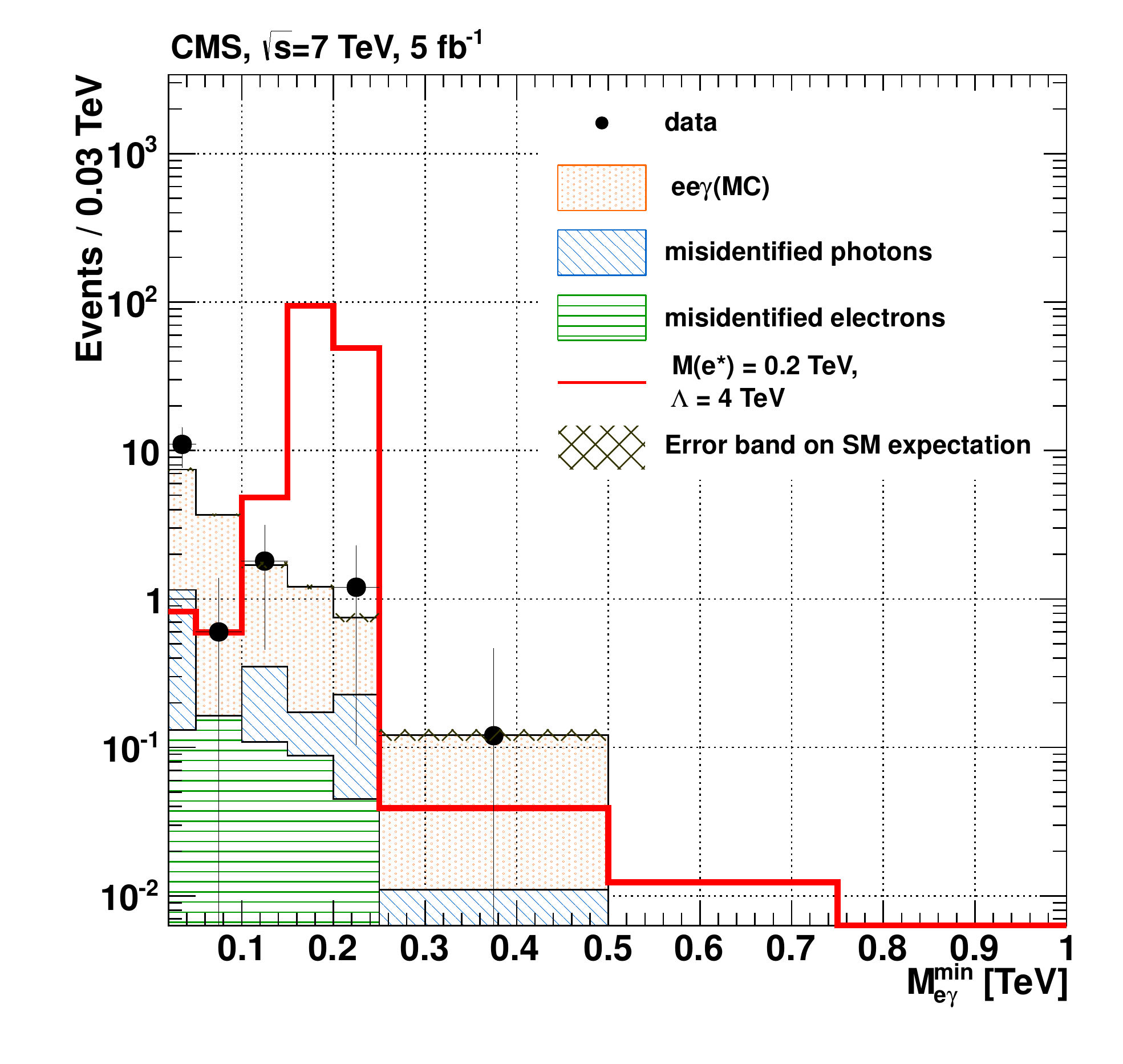}
\includegraphics[width=0.45\textwidth]{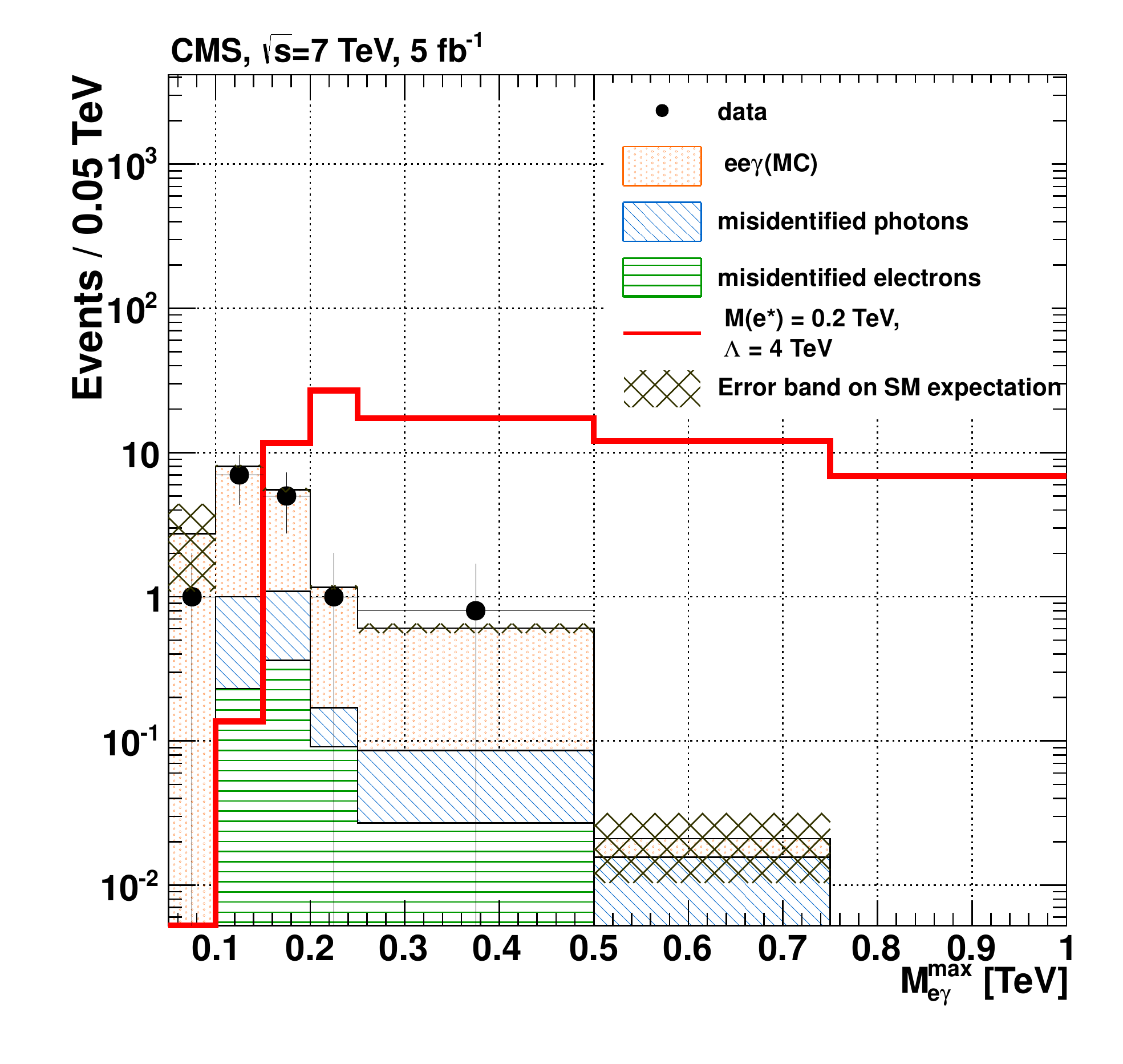}
\end{center}
\caption{The distribution of events as a function of \Mmin (\cmsLeft) and \Mmax (\cmsRight),
expected in the presence of an excited electron with a mass of 0.2\TeV.
The red dotted histogram corresponds to the contribution from
 the standard model backgrounds containing two real electrons and
a real photon.
The blue slanting hatched (green horizontal hatched)
 histograms correspond to the
contribution from misidentified photon (electrons).
The black solid circles correspond to the observed data.
 The red solid line histogram corresponds to the signal distribution for a mass of 0.2\TeV.
The dark grey double hatched region shows the uncertainty in the
SM expectation.
\label{fig:minmaxmass}}
\end{figure}

\begin{figure}[thbp]
\begin{center}
\includegraphics[width=0.45\textwidth]{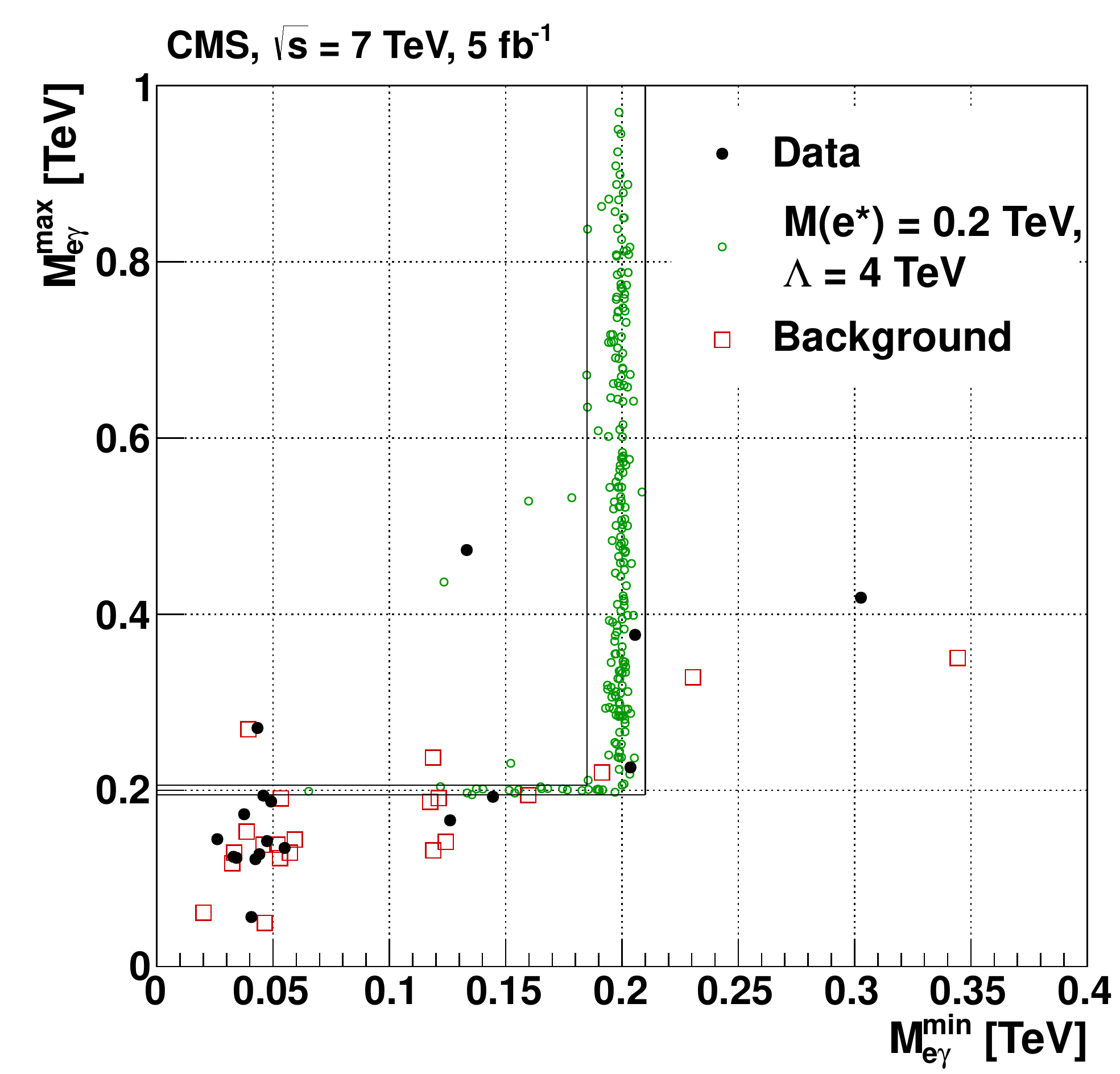}
\includegraphics[width=0.45\textwidth]{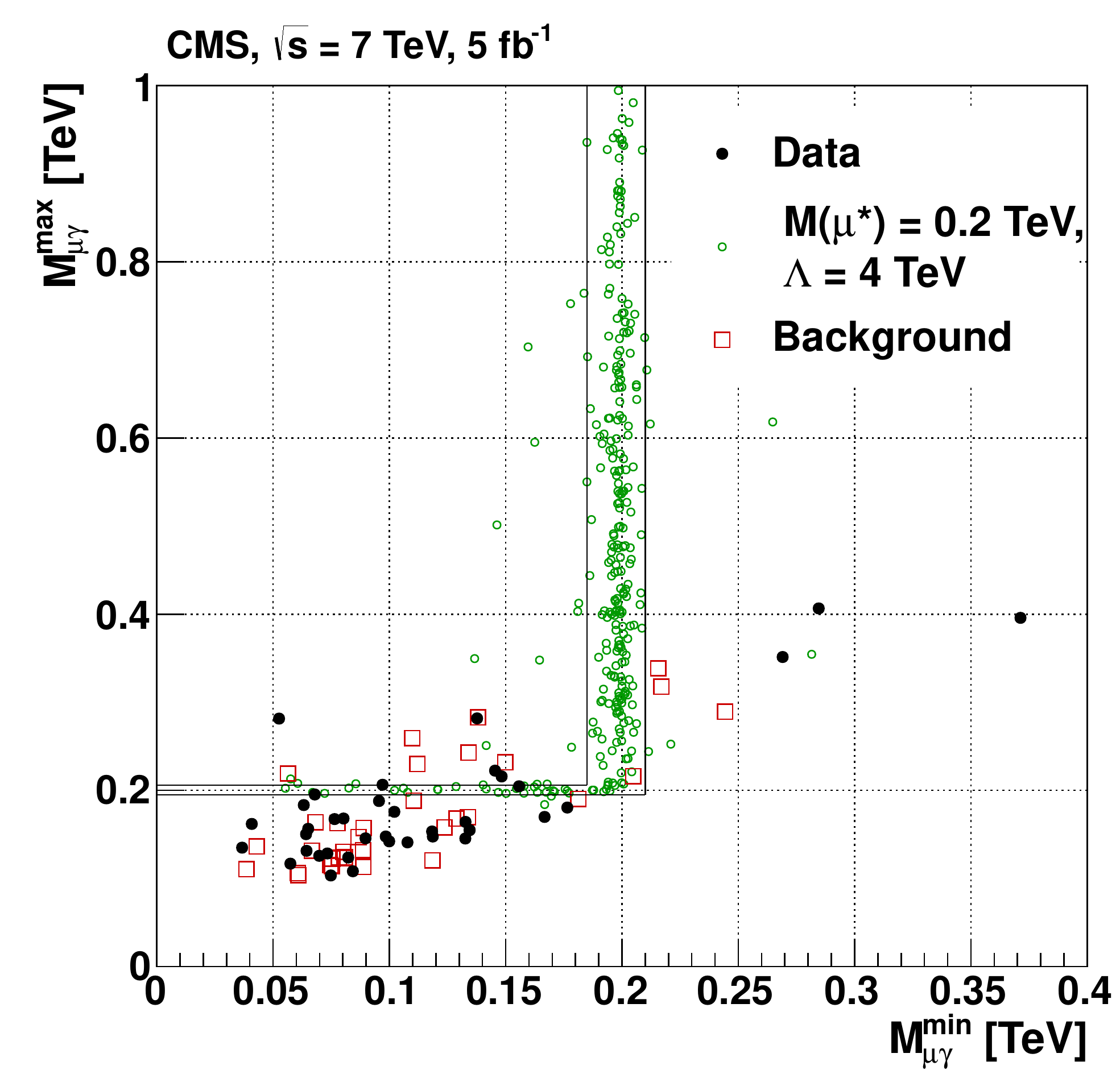}
\end{center}
\caption{Distribution of \Mmin and \Mmax for the excited electron analysis (\cmsLeft) and excited muon analysis (\cmsRight).
 The black solid circles, the red squares and the green open circles correspond to the observed data,
the background distribution and the signal distribution, respectively.
 The optimized selection boundaries are shown for an
excited lepton mass of 0.2\TeV. The sample is normalized to 5\fbinv of integrated luminosity.
\label{fig:2d}}
\end{figure}

Samples are
 obtained for different values of the
 excited lepton mass and $\Lambda$ = 4\TeV, with the CTEQ6L1~\cite{cteq} parametrization
for the parton distribution functions.
This particular choice of the value of $\Lambda$ has no impact on the simulated kinematics and all
results are presented independently of the value of
$\Lambda$, except for the signal yield
in Fig.~\ref{fig:minmaxmass} and Fig.~\ref{fig:2d}.
The SM background samples: $\cPZ + \gamma$, $\PW + \gamma$,  $\ttbar$, $\cPZ + \text{jets}$, $\PW + \text{jets}$, and $\PW\PW$ are generated with
{\MADGRAPH 4.5.1}~\cite{madgraph}. {\PYTHIA} has been used to perform the fragmentation and hadronization of samples
generated with {\MADGRAPH}.
The diboson samples (WZ, ZZ) are generated using {\PYTHIA 6.424}.
 The main background $\cPZ + \gamma$ has been generated to correspond to an integrated luminosity of around $7$\fbinv.
For all these SM background processes, the cross sections are scaled to
the next-to-leading order (NLO) cross sections obtained from
the parton level integrator \textsc{mcfm}~\cite{MCFM}.
 For the main background $\cPZ + \gamma$, the theoretical scale uncertainty has been evaluated using \textsc{mcfm} to be $+2.4$\%, $-1.6$\%.
 All Monte Carlo events used in this analysis have been passed through the detailed simulation of the CMS detector based on {\GEANTfour}~\cite{geant4}.

\section{Event reconstruction and selection }
\label{sec:selection}

Candidate events  for the electron  (muon) channel are  selected using triggers
with the lowest
 possible thresholds on lepton
transverse  momentum.   This  corresponds  to  a  transverse  momentum
threshold of  33~(24)\GeV for the initial periods  and 33~(40)\GeV for
the later periods of data
collection in the electron (muon) channel.
The trigger thresholds were raised in response to the increased mean instantaneous luminosity.
For the leptons selected in  the analysis,  the trigger  efficiencies are
100\%~(97\%)  in the electron  (muon)  channel.
The two leptons and the photon in signal events are expected to be
isolated from other particles in the event. This can be quantified by isolation
variables, obtained by summing the energy deposits present inside a
geometrical cone around the particle, in the tracker or in the calorimeters.
Events with at least one
well-reconstructed  primary vertex,
one isolated high-\pt photon, and two isolated high-\pt leptons are used in this analysis.

Electron  identification  is performed using
 clusters of localized energy deposits in the ECAL.
 An energy deposit in the
ECAL due to an electron is identified by imposing requirements on
shower shapes of the ECAL clusters and  isolation variables
as well as
the ratio of the energies deposited in the hadron and electromagnetic calorimeters
(${H/E}$).
A reconstructed track correctly associated with an ECAL cluster is
also required. For the electron channel, the electrons are required to
have  a  transverse  energy  $\et>35$~(40)\GeV in  the  ECAL  barrel
(endcap)  and $|\eta|<2.5$, excluding  the transition  region $1.4442<
|\eta|< 1.560$ between the ECAL barrel and endcap regions.
The  electron is  required  to be  isolated  both in  the tracker  and
calorimeter  within a cone  of radius  $\Delta R  \equiv \sqrt{(\Delta
\phi)^2 + (\Delta \eta)^2} < 0.3$ around its direction.
 In the tracker, the scalar sum of the \pt of the tracks, that are at least 0.7\GeV in \pt and lie outside a cone of
radius $\Delta R = 0.04$ relative to the electron, is required to be less than 5\GeV.
For the isolation using the calorimeters,
a variable \etiso is introduced, defined as
the total sum of transverse
energy deposits excluding  deposits  associated with  the  electron. In the
barrel, \etiso is
required to be less than $0.03 \et + 2.0$\GeV, and in the endcap:
 for $\et < 50$\GeV, the total \etiso is required to be below ${2.5}\GeV$; for $\et > 50$\GeV, it is required to be
 below $0.03\et + 1.0$\GeV.

For photons, identification criteria on the shower shapes, isolation variables and \textit{H/E} are
applied to energy clusters in the ECAL~\newbb \cite{PhoRec}\newbe. Photon candidates are
required to have clusters with $\et> 35\GeV$ and to be
in the central region (barrel) of the ECAL with $|\eta|<$1.4442.
The photon is also required to be isolated within a cone of radius $\Delta R < 0.4$ around its direction, both in the tracker and calorimeter.
\newbb The cone axis is taken to be the direction of the line joining the barycenter of the energy cluster
to the primary vertex. \newbe
In the tracker, the \newbb scalar \newbe sum of the transverse momenta of the tracks,  excluding tracks within an inner cone of
0.04, is required to be less  than $0.001 \pt + 2$\GeV.
In the ECAL, the total \etiso in the barrel, excluding deposits associated
with the photon, is required to be below $0.006 \et + 4.2$\GeV, whereas
for the HCAL isolation,
it is required to be below $0.0025 \et + 2.2$\GeV.

Muons
are reconstructed by combining tracks from the inner tracker and the outer muon system,
requiring at least one hit in the pixel tracker,
hits in more than 8 tracker layers and track segments reconstructed in at least two muon stations.
Since the segments have
multiple hits that typically occur in different
muon detectors and are therefore
separated by thick layers of iron, the latter requirement significantly
reduces the probability of a hadron being misidentified as a muon.
For the muon channel,
 two muons are required with each having  $|\eta|<$ 2.1; and the higher (lower) momentum muon must have
$\pt> 45$~(40)\GeV.
In order to reduce the cosmic-rays muon background,
the transverse impact parameters of both
muon tracks with respect to the
primary vertex of the event are required to be
less than 0.2\unit{cm} \newb and \newe muon pairs \newb that \newe are back-to-back in the
transverse plane are rejected, with the angle between two muon tracks \newb below \newe $\pi -0.02$.
Furthermore, the muon is required to be isolated such that the scalar  sum of the transverse momenta of all
tracks originating at the  interaction vertex, excluding the muon itself, within a $\Delta R < 0.3$ cone around
its direction is less than 10\% of its \pt.

In order to reject Drell--Yan events with final state radiation,
the distance  in ($\eta$, $\phi$) coordinates between the photon and the leading lepton,
$\Delta R(\ell, \gamma)$ is required to be $\Delta R(\ell, \gamma) > 0.5$ for $\ell=\mathrm{e}$
and $\Delta R(\ell, \gamma) > 0.7$ for $\ell=\mu$.
Two lepton-photon invariant masses can also be computed, because the final state is composed of two leptons and one photon.
For the electron channel, the dielectron invariant mass
is required to be above 60\GeV
and each of the dielectron and electron-photon invariant masses are required to be outside a $\pm25$\GeV window
centered at the nominal Z mass (91.19\GeV).
For the muon channel, the dilepton invariant mass is required to be 25\GeV above the nominal Z mass.
Fig.~\ref{fig:minmaxmass} shows the distribution of  \Mmin and \Mmax,
 the lower and higher invariant mass respectively.
 In the case of a signal, the correct assignment peaks at the excited lepton mass.
In the \Mmin-\Mmax plane, the signal is distributed along two
mutually perpendicular narrow bands. This shape determines the final
selection cuts as outlined below and is illustrated in Fig.~\ref{fig:2d} for $M_{\ell^*} = 0.2$\TeV. Identical boundaries are used for the electron and muon
channel. The only difference in the selection  between the two channels is the Z  veto, which, in the electron channel,
is also applied on  electron-photon
invariant mass.

 The background is located in the low invariant mass region,
while the signal populates the higher invariant mass region.
Using simulations, the boundaries of the signal region for a given mass have been chosen to
optimize the expected limit. The final values for different excited
lepton masses are shown in Table~\ref{table_inputnumbersummary}.
For $M_{\ell^*} = 0.2$\TeV, the horizontal band is small,
in order to reduce the background contamination.
For $M_{\ell^*} = 0.4$\TeV, a larger horizontal band can be used,
the increase of the background contamination being compensated by
the gain in signal efficiency.
For higher excited lepton masses,
the horizontal band is large to improve the signal efficiency in regions where almost no background is present.

\begin{table*}[htb]
\begin{center}
\topcaption{
Measured signal and expected background event numbers for the electron and muon channels as a function of
the mass of the excited lepton.
The signal efficiency with its corresponding uncertainty is given as
$\epsilon_{\textrm{signal}}$. The
expected numbers of background events
are reported as $N_{\textrm{bkgd}}$
with Clopper--Pearson errors~\cite{clopperpearson}
along with the observed data
$N_{\textrm{data}}$.
The boundaries values for \Mmin and \Mmax, which correspond to the signal region, are also given. The signal
efficiencies shown with $\dagger$ symbol are obtained from a polynomial curve fitted to the signal efficiencies
\newb for the mass points that have been simulated. \newe}
\renewcommand{\arraystretch}{1.7}
\begin{tabular}{lcc|ccc|ccc}
\hline
$M_{\ell^*}$ & \Mmin & \Mmax & \multicolumn{3}{c|}{Electron channel} & \multicolumn{3}{c}{Muon channel} \\
(TeV) & (TeV) & (TeV) & $\epsilon_{\textrm{signal}}$ (\%) & $N_{\textrm{bkgd}}$  & $N_{\textrm{data}}$
      & $\epsilon_{\textrm{signal}}$ (\%) & $N_{\textrm{bkgd}}$  & $N_{\textrm{data}}$ \\
\hline
0.2  & 0.19-0.21 &0.20-0.21 & 24.8 $\pm$ 1.8\phantomdagger & 1.0 $^{+1.1}_{-0.5}$ & 2 & 28.2 $\pm$ 1.3\phantomdagger & 1.2$^{+1.7}_{-0.6}$ & 2  \\
0.3  & 0.23-0.37 &0.29-0.31 & 30.0 $\pm$ 2.2\,$^{\dagger}$ & 1.2 $^{+2.1}_{-0.8}$ & 1 & 34.4 $\pm$ 1.6\,$^{\dagger}$ & 5.4$^{+2.6}_{-1.8}$ & 2  \\
0.4  & 0.28-0.52 &0.38-0.41 & 32.7 $\pm$ 2.4\phantomdagger & 0.1 $^{+1.4}_{-0.1}$ & 1 & 39.1 $\pm$ 1.8\phantomdagger & 1.6$^{+2.0}_{-0.9}$ & 3 \\
0.5  & 0.35-0.65 &0.47-0.53 & 34.8 $\pm$ 2.6\,$^{\dagger}$ & 0.0 $^{+1.4}_{-0.0}$ & 1 & 42.1 $\pm$ 1.9\,$^{\dagger}$ & 0.0$^{+1.4}_{-0.0}$ & 1  \\
0.6  & 0.42-0.78 &0.55-0.64 & 36.6 $\pm$ 2.6\phantomdagger & 0.0 $^{+1.4}_{-0.0}$ & 0 & 45.4 $\pm$ 2.0\phantomdagger & 0.0$^{+1.4}_{-0.0}$ & 0 \\
0.7  & 0.49-0.91 &0.65-0.76 & 37.8 $\pm$ 2.7\,$^{\dagger}$ & 0.1 $^{+1.4}_{-0.0}$ & 0 & 45.9 $\pm$ 2.1\,$^{\dagger}$ & 1.0$^{+1.7}_{-0.6}$ & 0  \\
0.8  & 0.56-1.04 &0.75-0.88 & 37.8 $\pm$ 2.7\phantomdagger & 0.0 $^{+1.4}_{-0.0}$ & 0 & 45.3 $\pm$ 2.0\phantomdagger & 0.0$^{+1.4}_{-0.0}$ & 0 \\
1.0  & 0.70-1.30 &0.75-1.08 & 40.4 $\pm$ 2.8\phantomdagger & 0.0 $^{+1.4}_{-0.0}$ & 0 & 48.5 $\pm$ 2.1\phantomdagger & 0.0$^{+1.4}_{-0.0}$ & 0 \\
1.2  & 0.84-1.56 &0.75-1.34 & 41.1 $\pm$ 2.9\phantomdagger & 0.0 $^{+1.4}_{-0.0}$ & 0 & 50.0 $\pm$ 2.2\phantomdagger & 0.0$^{+1.4}_{-0.0}$ & 0 \\
1.5  & 1.05-1.95 &0.75-1.67 & 41.7 $\pm$ 2.9\phantomdagger & 0.0 $^{+1.4}_{-0.0}$ & 0 & 50.8 $\pm$ 2.2\phantomdagger & 0.0$^{+1.4}_{-0.0}$ & 0 \\
2.0  & 1.40-2.60 &0.75-2.23 & 43.5 $\pm$ 3.1\phantomdagger & 0.0 $^{+1.4}_{-0.0}$ & 0 & 50.4 $\pm$ 2.2\phantomdagger & 0.0$^{+1.4}_{-0.0}$ & 0 \\
\hline
\end{tabular}
\label{table_inputnumbersummary}
\end{center}
\end{table*}

\section{Background due to particle misidentification}
\label{sec:bkg}

\begin{figure}[h!tbp]
\centering
\includegraphics[width=\cmsFigWidth]{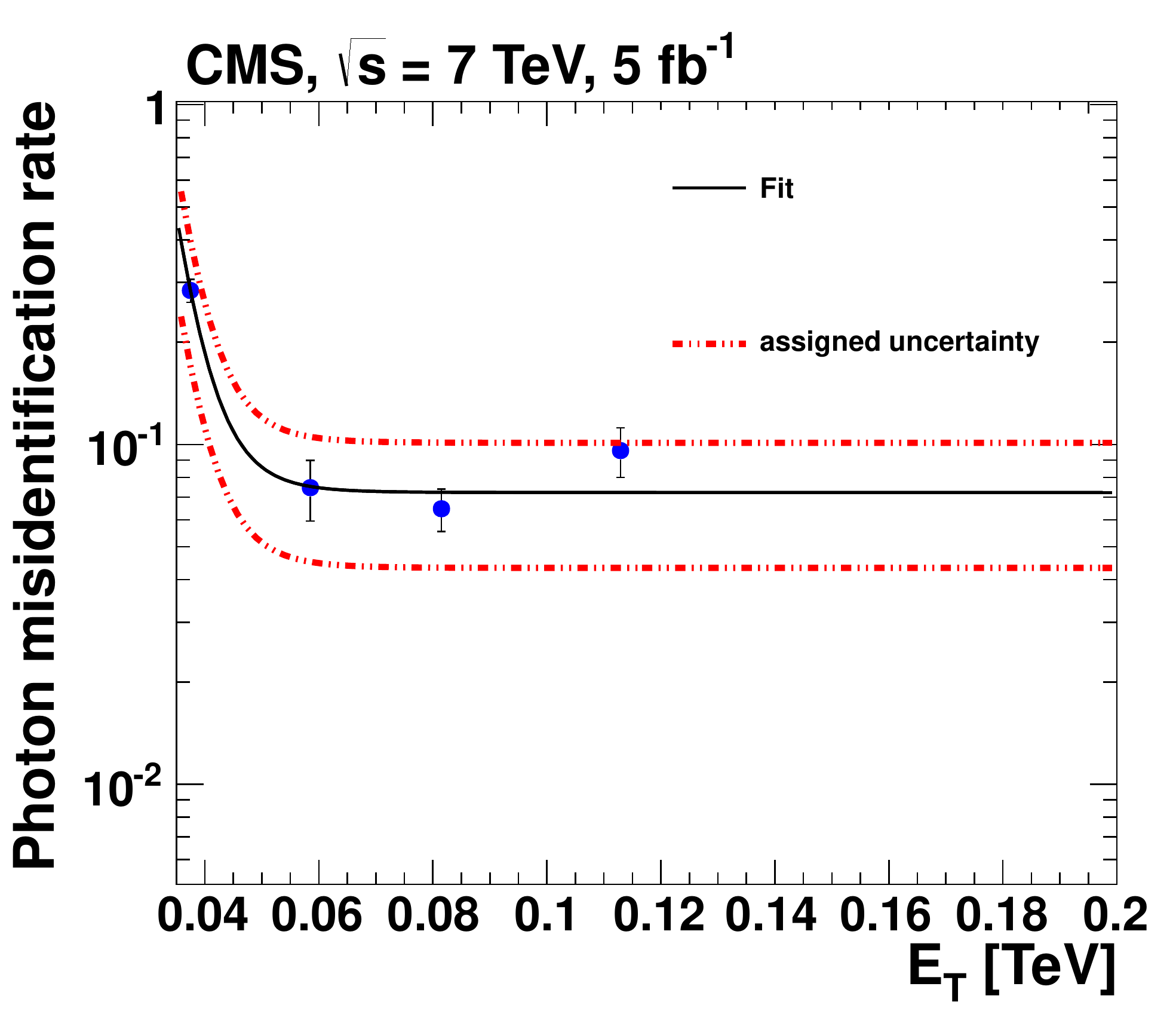}
\caption{The jet-to-photon misidentification rate as a function of \et. The dashed line is the 40\% uncertainty band.
\label{fig:fakerateVspT}
}
\end{figure}

Hadronic jets in which a $\pi^{0}$  carries a significant fraction of the energy
may be misidentified as isolated photons.
Thus Z+jets events are a potential background for this search.
The photon misidentification rate is
measured directly from
a data sample dominated by jets, with a photon-like candidate
cluster embedded inside, which can potentially
be misidentified as a photon. The misidentification rate
is defined as the ratio of the number of
photon candidates passing all the photon
selection criteria (numerator) to the  number of photon
candidates that pass a loose set of shower shape requirements
but fail one of the photon isolation criteria (denominator).
The misidentification rate is estimated in bins of photon ${\et}$.
The numerator sample can have a contribution from
isolated true photons.
\newb This misidentification rate is therefore corrected by using the probability distribution of
energy-weighted shower width ($\sigma_{\eta \eta}$) of isolated true photons
computed in units of crystal size, which is different from that of non-isolated photons. \newe
The true photon fraction in the numerator is estimated by
fitting these two different shower shapes to the
shower shape distribution of the numerator sample,
and subtracted from the numerator. In order to estimate the contribution of misidentified photons in the
analysis, \newb the \newe misidentification rate is applied to a subsample of data events containing one
photon candidate and satisfying all other selection criteria. This rate is calculated in photon \et
bins of (0.03--0.05, 0.05--0.075, 0.075--0.09, 0.09--0.2)\TeV. Fig.~\ref{fig:fakerateVspT} shows the \et dependence of the
photon misidentification rate.
The calculated misidentified photon rate is found to be 0.28, 0.07, 0.06 and
0.09 for the above mentioned \et bins.

From a fit, the measured rate is parametrized by a function, $f_{\gamma}^{\text{misid}}(\ET)$, as given in the equation (\ref{eqn:fr})
 with $a$, $b$ and $c$ being the fit parameters:

\begin{equation}
\label{eqn:fr}
f_{\gamma}^{\text{misid}}(\ET) = a + \frac{b}{{(\ET)^c}}.
\end{equation}

An uncertainty
of 40\% is assigned to this function which envelopes the spread of data points relative to the fit.
\newb The jet to photon misidentication is estimated by applying this misidentification rate
to a sample passing all our selection requirements, including triggers, except a requirement that the photon candidate
fails one of the photon identification criteria and passes instead the loose identification requirements.
Applied to the lowest mass point of 0.2\TeV, \newe
the contribution of photon misidentification background in the full selection is
found to be 0.07$^{+0.16}_{-0.07}$ events for
both the electron
and the muon channels.
It is negligible for higher mass points.

\begin{table*}[htb]
\begin{center}
\renewcommand{\arraystretch}{1.7}
\topcaption{Details of the expected background compositions for several masses, showing contributions from $\cPZ + \gamma$ MC sample, misidentified $\gamma$ and
misidentified electron estimated from data.
The uncertainties are reported as the quadratic sum of statistical and systematic errors.}
\begin{tabular}{l|ccc|cc}
\hline
$M_{\ell^*}$  & \multicolumn{3}{c}{Electron channel} & \multicolumn{2}{|c}{Muon channel} \\
(TeV) &  $\cPZ + \gamma$ MC & misid $\gamma$  & misid electron
      &  $\cPZ + \gamma$ MC & misid $\gamma$ \\
\hline
 0.2 & 0.8$^{+1.1}_{-0.5}$ & 0.07$^{+0.16}_{-0.07}$& 0.08$^{+0.17}_{-0.07}$
 & 1.0 $^{+1.7}_{-0.6}$ & 0.07 $^{+0.16}_{-0.07}$ \\
 0.4 & 0.0$^{+1.4}_{-0.0}$ & 0.07$^{+0.16}_{-0.07}$ & 0.01$^{+0.02}_{-0.01}$
 & 1.6 $^{+1.9}_{-0.9}$ & 0.00 $^{+0.45}_{-0.00}$ \\
$\geq$0.6 & 0.0$^{+1.4}_{-0.0}$ & 0.00$^{+0.45}_{-0.00}$  & 0.00$^{+0.08}_{-0.00}$
 & 0.0 $^{+1.4}_{-0.0}$ & 0.00 $^{+0.45}_{-0.00}$ \\
\hline
\end{tabular}
\label{table_bkgdsumm}
\end{center}
\end{table*}

Backgrounds with zero or one real electron can contribute to the \estar search.
The largest contributions come from processes such as
$\PW (\rightarrow {\Pe}\nu)$ + jet + $\gamma$ where the jet in the event is
misidentified as an electron.
Misidentification can occur when
photons coming from $\pi^{0}$s inside a jet convert to an \Pep\Pem\  pair
and are misidentified as electrons. Other possible sources include when a
charged particle within a jet provides both the track in the tracker and
an electromagnetic cluster
that together fake an electron signature,
or when a track from a charged particle matches with a nearby energy deposition
in the calorimeter from another particle.
The misidentification rate is calculated as the ratio between
the number of candidates passing the electron selection criteria with respect to those
 satisfying looser selection criteria.
The looser selection criteria require only that the first
tracker layer contributes a hit to the
electron track and that
offline emulations
of the online trigger requirements (``loose identification requirements'') on shower
shape $\sigma_{\eta \eta}$ and the ratio ${H/E}$ are satisfied.
This ratio is estimated as a function of
\et in bins of $\eta$ ($f_\text{electron}^{\text{misid}}(\ET, \eta)$) using a
data sample selected with single-photon triggers~\cite{zprimetoee}.
The jet to electron misidentified background in \estar is estimated
 by applying this misidentification rate to a sample passing all
our selection requirements, including triggers,
except requiring one of the electron candidates to fail the
electron identification criteria and pass instead the loose identification requirements.
The systematic uncertainty on $f_\text{electron}^{\text{misid}}(\ET, \eta)$
 is determined using a sample of events containing two reconstructed
electrons as in~\cite{zprimetoee}.
The  contribution from jet events to the dielectron mass spectrum
can be determined
either by applying the misidentification rate twice on events with two loose electrons
or by applying the misidentification rate once on events with one fully identified electron and one
loose electron. The first estimate lacks contributions from W + jets and $\gamma$ + jets
 events while the second estimate is contaminated
by Drell--Yan events. These effects are corrected using simulated samples.
If the misidentification rate method is correct, the two corrected estimations should agree.
Both estimates are found to agree well and the residual difference of 40\% \newb between the two estimates \newe is taken
as \newb the \newe systematic uncertainty on the jet to electron misidentification rate.
The contribution from events which have zero or one real electron
is $0.08^{+0.17}_{-0.07}$ for the lowest mass point of 0.2\TeV and is negligible for higher mass points.

\section{Results}
\label{sec:results}

After all selection steps the expected background for $M_{\ell^*}>0.7$\TeV is found to be
$0^{+1.4}_{-0.0}$ event in the simulated sample.
The signal efficiency increases with the mass of the excited lepton, from 25\% to 44\% in the electron channel
and 28\% to 50\% in the muon channel.
All numbers are summarized in Table~\ref{table_inputnumbersummary}.
The expected numbers of signal events and irreducible background events
are evaluated from simulation while the contribution of misidentified  particles is derived from data. The background composition for several  mass points, 0.2\TeV, 0.4\TeV and $\geq$ 0.6\TeV for both channels is shown in Table~\ref{table_bkgdsumm}.
The uncertainties in the description of the detector performance,
such as lepton energy or momentum resolution, lepton and photon energy  scales, have been included in the systematic uncertainties.
The impact on the signal yield
corresponds to an uncertainty of $\pm$2\% and $\pm$3.5\%, for \newb the electron and muon channels respectively. \newe
Effects caused by the increase in the typical number of additional pp  interactions (`pileup') per LHC bunch crossing are modeled by adding to  the generated events multiple
collisions
with a multiplicity distribution matched to the luminosity profile of the collision data. To evaluate the
systematic uncertainty associated with the pileup simulation, the mean of the distribution of the pileup
interactions is varied by 5\%, leading to a variation of 3.0\%~(0.6\%) in the simulated backgrounds and
1.0\%~(1.5\%) in signal yields in the electron (muon) channel.
An additional systematic uncertainty of 10\%
is assigned to the background to account for uncertainties associated with the choice of
parton distribution functions. The uncertainty in the luminosity normalization is 2.2\%~\cite{lumi}.

\begin{figure}[thbp]
\begin{center}
\includegraphics[width=0.45\textwidth]{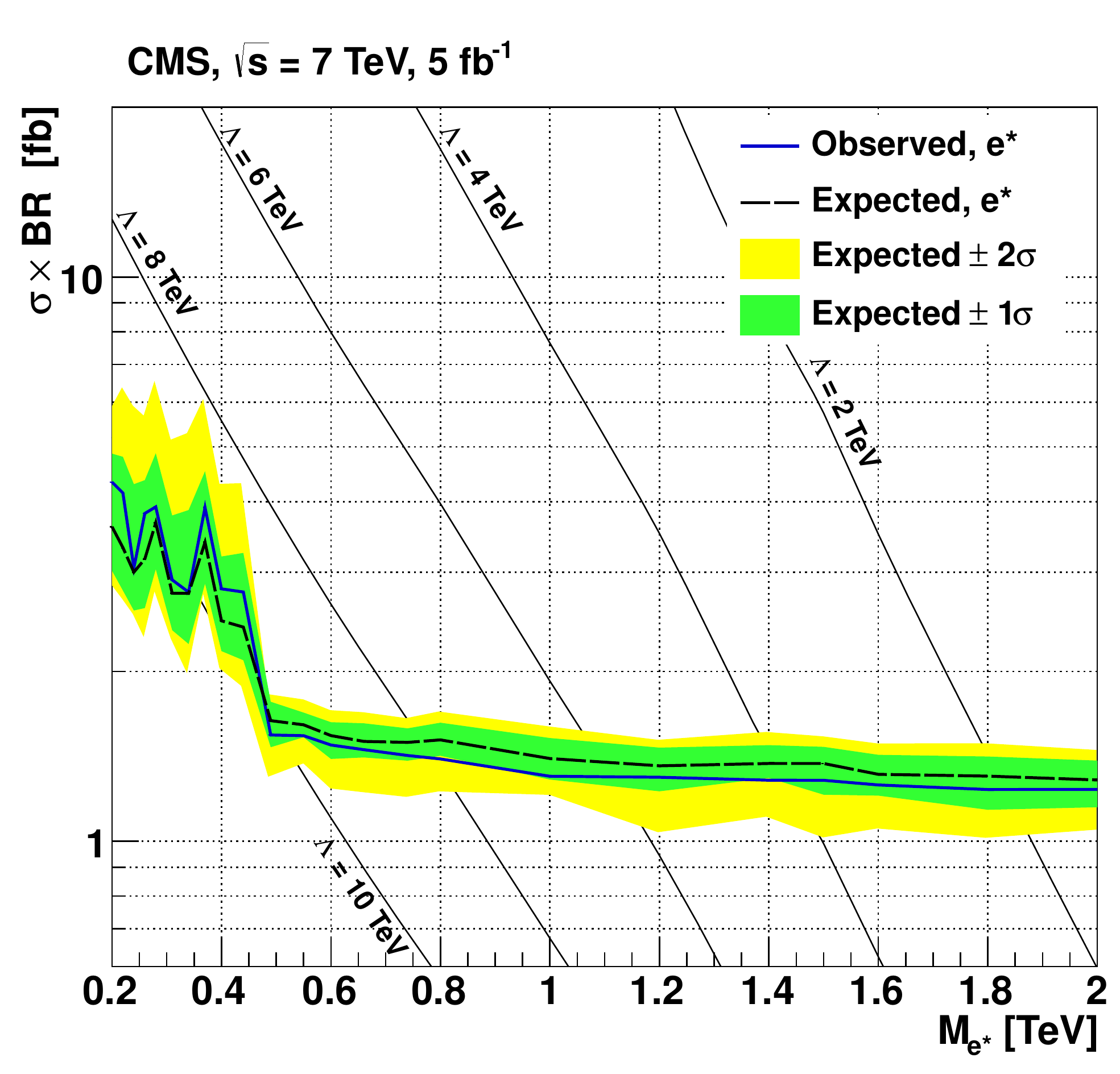}
\includegraphics[width=0.45\textwidth]{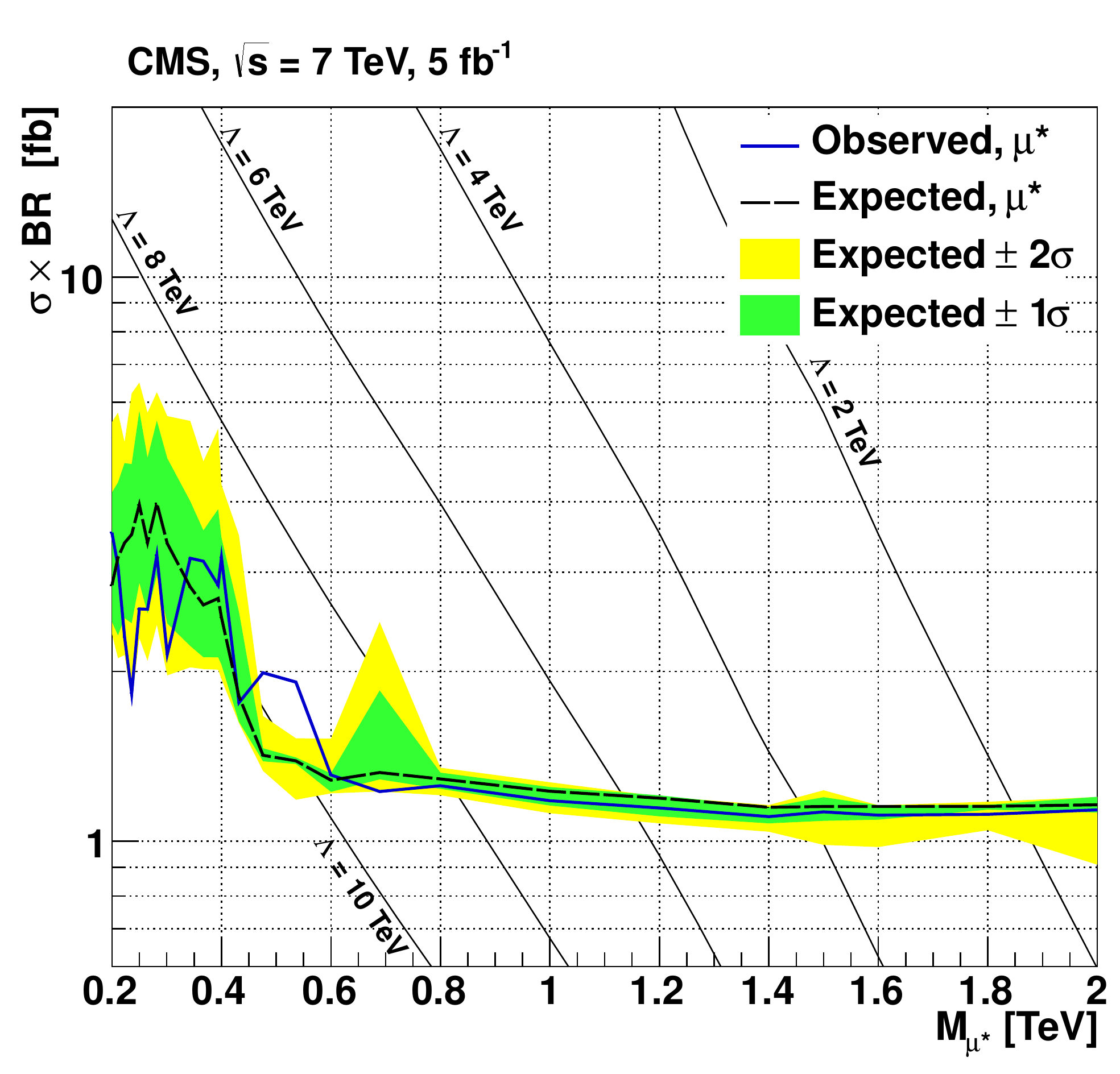}
\end{center}
\caption{Expected and observed 95\% CL upper limits on the cross section of the studied channel for the
different excited electron (\cmsLeft) and muon (\cmsRight)
mass points,
 using the $\textrm{CL}_{\textrm{s}}$ method.
\newb The excluded region is above the curve. \newe
The black solid lines correspond to the excited lepton LO cross sections times branching ratio for
different $\Lambda$
scales. The one (two) standard deviation uncertainty
 bands are shown in green
(yellow).
\label{fig:limit}
}
\end{figure}

As seen in Table~\ref{table_inputnumbersummary}, for masses above 0.5\TeV,
no data events pass
the criteria designed to select excited lepton signatures.
 Using a single bin counting method, upper limits are provided on the production
cross section
times branching fraction of excited electrons
and excited muons at the 95\%  confidence level.
The method is implemented in the statistical package developed by the Higgs study
group~\cite{higgstool}.
The computation has been performed using both a Bayesian~\cite{Bayes,PDGBayes}
and a $\mathrm{CL}_\mathrm{s}$~\cite{CLs,Junk:1999kv} approach; the results are found to be consistent with each
other.
The results presented here are from the frequentist CL$_\mathrm{s}$ approach, without the use of the asymptotic
approximation~\cite{higgstool}.
The background and signal uncertainties are dominated by completely uncorrelated uncertainties. The integrated luminosity
normalization uncertainty is
considered separately,
with 100\% correlation between signal and background.
The nuisance parameters related to the uncertainties on the background  are treated according to gamma
probability distribution functions.
The uncertainties on the signal yield and the integrated luminosity  normalization are taken into account via a
lognormal treatment of  nuisance parameters.
The observed limits for the electron and the muon channels are shown in Fig.~\ref{fig:limit}.
Production cross sections higher than 1.48 to 1.24\unit{fb} (1.31 to 1.11\unit{fb})
are excluded at the 95\% confidence limit (CL)
for \estar (\mustar) masses ranging from 0.6 to 2\TeV.
The structure observed in the expected and observed limits results from the limited
sizes of the simulated background samples.
The optimization of the invariant masses selecting the \Mmin--\Mmax signal region has been determined from
simulation of signal
reference mass points, ranging from $M_{\ell^*}$ = 0.2\TeV to 2.0\TeV in steps of 0.2\TeV. For
lower masses, the selected signal regions do not overlap.
For continuous coverage, additional mass points for $M_{\ell^*} < 0.6\TeV$, have been added by interpolating
the cut thresholds
and the signal efficiencies.
Limits for masses between 0.2 and 0.4\TeV are less stringent because of the presence of background in this region.

In the excited muon channel, as visible in Table~\ref{table_inputnumbersummary}, the bump at $M_{\mu^*} \sim$ 0.5\TeV corresponds
to a region where the background is found to be $0.0^{+1.4}_{-0.0}$ in the simulated sample while one data event
is observed.
Also in this channel, the shape of the uncertainty bands at $M_{\mu^*} = 0.7\TeV$ corresponds to a region where the background is
found to be $1.0^{+1.7}_{-0.6}$ in the simulated sample while zero data events are observed.
\newbb
For high excited lepton masses, the muon channel cross section limit is slightly lower  than the
electron channel limit because of the difference in the acceptance. For lower excited lepton masses,
the sensitivity of the electron channel is also reduced because of misidentification of photons and
electrons.
\newbe

\begin{figure}[thbp]
\begin{center}
\includegraphics[width=0.45\textwidth]{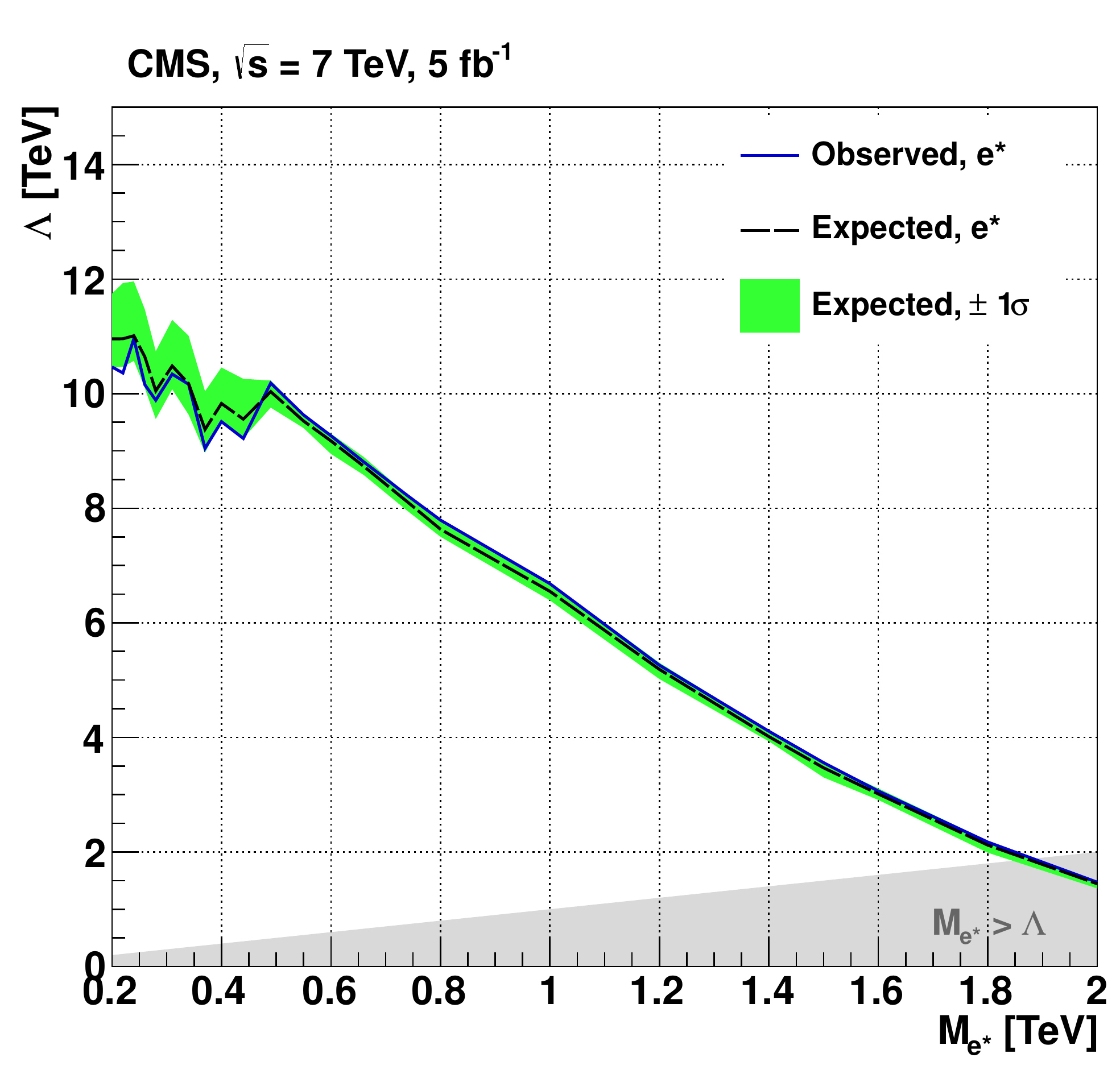}
\includegraphics[width=0.45\textwidth]{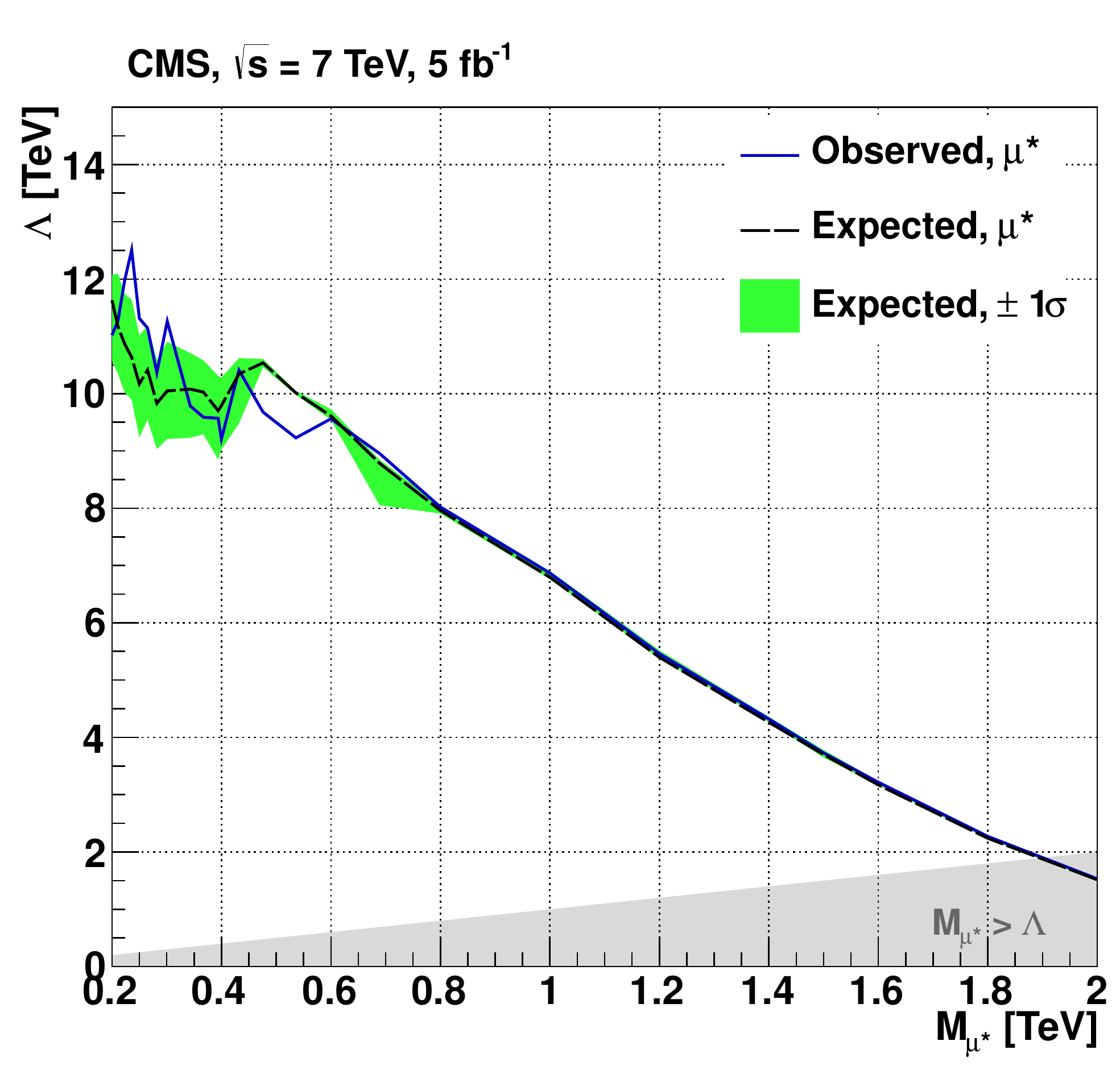}
\end{center}
\caption{Expected and observed 95\% CL lower limits on the $\Lambda$ scale for the different excited
electron (\cmsLeft)
and muon (\cmsRight) mass points,
 using the $\textrm{CL}_{\textrm{s}}$ method.
\newb The excluded region is below the curve. \newe
These limits are computed with the LO signal cross section obtained from {\PYTHIA 6.424}. The one standard deviation
uncertainty band is shown in green. The bands do not include the uncertainty on signal cross section. \newb The grey area corresponds to the theoretically excluded region where $M_{\ell^*} > \Lambda$.\newe
\label{fig:lambda}
}
\end{figure}

The set of $\Lambda-M_{\ell^*}$ values for which the theoretical cross section times branching fraction
 is higher than the 95\% upper limit on cross section, is
considered as excluded region of the parameter space.
 The exclusion region in the $\Lambda-M_{\ell^*}$ plane is shown in Fig.~\ref{fig:lambda}.
The displayed uncertainty band corresponds to the uncertainty
on the cross section limits, and does not take into account uncertainties
on the theoretical signal cross section.
\newb The region is theoretically excluded, where $M_{\ell^*} > \Lambda$. \newe
 The signal cross sections are estimated
with the $Q^{2}$ scale set to the square of the mass of excited lepton ($M_{\ell^*}^{2}$).
If the $Q^{2}$ scale is varied
to $M_{\ell^*}^{2}$/2, the limit for $\Lambda = M_{\ell^*}$ increases by 1.5\% and
if it is varied to 2$M_{\ell^*}^{2}$, the limit for $\Lambda = M_{\ell^*}$  decreases by 2.4\%.
 \newb The impact of the parton distribution functions (PDF) uncertainties on the signal is smaller than 1\%. \newe

Assuming the same masses for \estar and \mustar, the two counting experiments have been combined using the
CL$_\textrm{s}$ approach, improving the excluded cross section limit to
0.73 to 0.60\unit{fb} for masses from 0.6 to 2\TeV. \newb Allowing \estar and \mustar to have different masses,
the excluded cross sections would also be within this range. \newe
 The following uncertainties have been considered as completely
correlated between the two channels:  the photon scale factor  uncertainties in signal and background, the
photon misidentification rate systematic  uncertainty not related to statistics, the luminosity uncertainty,
the  pileup simulation uncertainty, the $\cPZ + \gamma$ normalization uncertainty, and the $\cPZ + \gamma$ PDF
uncertainty. The other uncertainties are considered as 100\% uncorrelated.

\section{Summary}

A search has been performed with the CMS detector for excited leptons in the electron ($\mathrm{pp}
\rightarrow
\Pe\Pe^* \rightarrow \Pe\Pe\gamma$) and muon
($\Pp\Pp\rightarrow \mu\mu^* \rightarrow \mu\mu\gamma$)
channels.
For each excited lepton mass, the excluded cross section can be associated with a value for the new
interaction scale $\Lambda$.
Excited leptons (electrons or muons) with masses below  1.9\TeV
are excluded for the scale
of contact interaction $\Lambda = M_{\ell^*}$.
Production cross sections higher than
1.48 to 1.24\unit{fb} (1.31 to 1.11\unit{fb})
are excluded at the 95\% CL
for \estar (\mustar) masses ranging from 0.6 to 2\TeV.
\newbb
The slightly better sensitivity in the muon channel is due to its better acceptance and efficiency, and
also, for lower $\ell^*$  masses, to the fact that there is a higher background background in the electron
channel arising from particle misidentification.
\newbe
These limits are the most stringent published to date.

\section*{Acknowledgements}

We congratulate our colleagues in the CERN accelerator departments for the excellent performance of the LHC
machine. We thank the technical and
administrative staff at CERN and other CMS institutes, and acknowledge support from: FMSR (Austria); FNRS
and FWO (Belgium); CNPq, CAPES, FAPERJ, and
FAPESP (Brazil); MES (Bulgaria); CERN; CAS, MoST, and NSFC (China); COLCIENCIAS (Colombia); MSES (Croatia);
RPF (Cyprus); MoER, SF0690030s09 and ERDF
(Estonia); Academy of Finland, MEC, and HIP (Finland); CEA and CNRS/IN2P3 (France); BMBF, DFG, and HGF
(Germany); GSRT (Greece); OTKA and NKTH (Hungary);
DAE and DST (India); I$\pm$ (Iran); SFI (Ireland); INFN (Italy); NRF and WCU (Korea); LAS (Lithuania);
CINVESTAV, CONACYT, SEP, and UASLP-FAI (Mexico); MSI
(New Zealand); PAEC (Pakistan); MSHE and NSC (Poland); FCT (Portugal); JINR (Armenia, Belarus, Georgia,
Ukraine, Uzbekistan); MON, RosAtom, RAS and RFBR
(Russia); MSTD (Serbia); MICINN and CPAN (Spain); Swiss Funding Agencies (Switzerland); NSC (Taipei); TUBITAK
and TAEK (Turkey); STFC (United Kingdom); DOE
and NSF (USA).
Individuals have received support from the Marie-Curie programme and the European Research Council (European
Union); the Leventis Foundation; the A. P.
Sloan Foundation; the Alexander von Humboldt Foundation; the Belgian Federal Science Policy Office; the Fonds
pour la Formation \`a la Recherche dans
l'Industrie et dans l'Agriculture (FRIA-Belgium); the Agentschap voor Innovatie door Wetenschap en
Technologie (IWT-Belgium); the Council of Scientific and
Industrial Research, India; and the HOMING PLUS programme of Foundation for Polish Science, cofinanced from
European Union, Regional Development Fund.
\bibliography{auto_generated}   % will be created by the tdr script.

\providecommand{\href}[2]{#2}\begingroup\raggedright\begin{thebibliography}{10}%
\makeatletter
\providecommand{\hrefCMSnoop }[0]{\@secondoftwo}%
\makeatother
\providecommand{\doi}{\texttt{doi:}\begingroup \urlstyle{tt}\Url}

\bibitem{Pati:1975md}
\hrefCMSnoop {} {J.~C. Pati, A.~Salam, and J.~A. Strathdee, ``Are quarks
  composite?'',} \textit{ Phys. Lett. B} \textbf{ 59} (1975) 265,
\href{http://dx.doi.org/10.1016/0370-2693(75)90042-8}{\doi{10.1016/0370-2693(75)90042-8}}.
%%CITATION = PHLTA,B59,265;%%.

\bibitem{compositeness1}
\hrefCMSnoop {} {H.~Terazawa, M.~Yasu{\`e}, K.~Akama, and M.~Hayshi,
  ``Observable effects of the possible substructure of leptons and quarks'',}
  \textit{ Phys. Lett. B} \textbf{ 112} (1982) 387,
  \href{http://dx.doi.org/10.1016/0370-2693(82)91075-9}{\doi{10.1016/0370-2693(82)91075-9}}.

\bibitem{Eichten1982}
M.~Abolins\href {http://inspirehep.net/record/185862/files/C8206282-pg274.PDF}
  { {et~al.}, ``{Testing the Compositeness of Quarks and Leptons}'',} in
  \textit{ Elementary Particles and Future Facilities (Snowmass 1982)}, p.~274.
\newblock 1982.
\newblock
eConf C8206282.
%%CITATION = ECONF,C8206282,274;%%.

\bibitem{Eichten:1983hw}
\hrefCMSnoop {} {E.~Eichten, K.~D. Lane, and M.~E. Peskin, ``{New Tests for
  Quark and Lepton Substructure}'',} \textit{ Phys. Rev. Lett.} \textbf{ 50}
  (1983) 811,
\href{http://dx.doi.org/10.1103/PhysRevLett.50.811}{\doi{10.1103/PhysRevLett.50.811}}.
%%CITATION = PRLTA,50,811;%%.

\bibitem{harari}
\hrefCMSnoop {} {H.~Harari, ``Composite models for quarks and leptons'',}
  \textit{ Physics Reports} \textbf{ 104} (1984) 159,
\href{http://dx.doi.org/10.1016/0370-1573(84)90207-2}{\doi{10.1016/0370-1573(84)90207-2}}.
%%CITATION = ECONF,C8206282,274;%%.

\bibitem{ssc-physics}
\hrefCMSnoop {} {K.~D. Lane, F.~E. Paige, T.~Skwarnicki, and W.~J. Womersley,
  ``{Simulations of supercollider physics}'',} \textit{ Phys. Rept.} \textbf{
  278} (1997) 291,
  \href{http://dx.doi.org/10.1016/S0370-1573(96)00018-X}{\doi{10.1016/S0370-1573(96)00018-X}},
\href{http://www.arXiv.org/abs/hep-ph/9412280}{\texttt{ arXiv:hep-ph/9412280}}.
%%CITATION = HEP-PH/9412280;%%.

\bibitem{Baur90}
\hrefCMSnoop {} {U.~Baur, M.~Spira, and P.~M. Zerwas, ``Excited quark and
  lepton production at hadron colliders'',} \textit{ Phys. Rev. D} \textbf{ 42}
  (1990) 815,
\href{http://dx.doi.org/10.1103/PhysRevD.42.815}{\doi{10.1103/PhysRevD.42.815}}.
%%CITATION = PHRVA,D42,815;%%.

\bibitem{Greenberg:1974qb}
\hrefCMSnoop {} {O.~W. Greenberg and C.~A. Nelson, ``Composite Models of
  Leptons'',} \textit{ Phys. Rev. D} \textbf{ 10} (1974) 2567,
\href{http://dx.doi.org/10.1103/PhysRevD.10.2567}{\doi{10.1103/PhysRevD.10.2567}}.
%%CITATION = PHRVA,D10,2567;%%.

\bibitem{Greenberg:1980ri}
\hrefCMSnoop {} {O.~W. Greenberg and J.~Sucher, ``A quantum structure dynamic
  model of quarks, leptons, weak vector bosons, and {H}iggs mesons'',} \textit{
  Phys. Lett. B} \textbf{ 99} (1981) 339,
\href{http://dx.doi.org/10.1016/0370-2693(81)90113-1}{\doi{10.1016/0370-2693(81)90113-1}}.
%%CITATION = PHLTA,B99,339;%%.

\bibitem{Buskulic:1996tw}
\hrefCMSnoop {} {{ ALEPH} Collaboration, ``Search for excited leptons at
  130--140 {GeV}'',} \textit{ Phys. Lett. B} \textbf{ 385} (1996) 445,
  \href{http://dx.doi.org/10.1016/0370-2693(96)00961-6}{\doi{10.1016/0370-2693(96)00961-6}}.

\bibitem{Abreu:1998jw}
\hrefCMSnoop {} {{ DELPHI} Collaboration, ``Search for composite and exotic
  fermions at {LEP 2}'',} \textit{ Eur. Phys. J. C} \textbf{ 8} (1999) 41,
  \href{http://dx.doi.org/10.1007/s100529901074}{\doi{10.1007/s100529901074}},
  \href{http://www.arXiv.org/abs/hep-ex/9811005}{\texttt{
  arXiv:hep-ex/9811005}}.

\bibitem{Abbiendi:1999sa}
\hrefCMSnoop {} {{ OPAL} Collaboration, ``Search for unstable heavy and excited
  leptons at {LEP 2}'',} \textit{ Eur. Phys. J. C} \textbf{ 14} (2000) 73,
  \href{http://dx.doi.org/10.1007/s100520050734}{\doi{10.1007/s100520050734}},
  \href{http://www.arXiv.org/abs/hep-ex/0001056}{\texttt{
  arXiv:hep-ex/0001056}}.

\bibitem{Achard:2003hd}
\hrefCMSnoop {} {{ L3} Collaboration, ``Search for excited leptons at {LEP}'',}
  \textit{ Phys. Lett. B} \textbf{ 568} (2003) 23,
  \href{http://dx.doi.org/10.1016/j.physletb.2003.05.004}{\doi{10.1016/j.physletb.2003.05.004}},
  \href{http://www.arXiv.org/abs/hep-ex/0306016}{\texttt{
  arXiv:hep-ex/0306016}}.

\bibitem{H1estar}
\hrefCMSnoop {} {{ H1} Collaboration, ``Search for excited electrons in $ep$
  collisions at {HERA}'',} \textit{ Phys. Lett. B} \textbf{ 666} (2008) 131,
  \href{http://dx.doi.org/10.1016/j.physletb.2008.07.014}{\doi{10.1016/j.physletb.2008.07.014}},
  \href{http://www.arXiv.org/abs/0805.4530}{\texttt{ arXiv:0805.4530}}.

\bibitem{CDFestar}
\hrefCMSnoop {} {{ CDF} Collaboration, ``{Search for Excited and Exotic
  Electrons in the $e\gamma$ Decay Channel in $\rm {p}\overline{\rm p}$
  Collisions at $\sqrt{s}$ = 1.96 TeV}'',} \textit{ Phys. Rev. Lett.} \textbf{
  94} (2005) 101802,
  \href{http://dx.doi.org/10.1103/PhysRevLett.94.101802}{\doi{10.1103/PhysRevLett.94.101802}},
  \href{http://www.arXiv.org/abs/hep-ex/0410013}{\texttt{
  arXiv:hep-ex/0410013}}.

\bibitem{cdfmu}
\hrefCMSnoop {} {{ CDF} Collaboration, ``{Search for Excited and Exotic Muons
  in the $\mu\gamma$ Decay Channel in $\rm {p}\overline{\rm p}$ Collisions at
  $\sqrt{s}$ = 1.96 TeV}'',} \textit{ Phys. Rev. Lett.} \textbf{ 97} (2006)
  191802,
  \href{http://dx.doi.org/10.1103/PhysRevLett.97.191802}{\doi{10.1103/PhysRevLett.97.191802}},
  \href{http://www.arXiv.org/abs/hep-ex/0606043}{\texttt{
  arXiv:hep-ex/0606043}}.

\bibitem{d0}
\hrefCMSnoop {} {{ D0} Collaboration, ``{Search for excited muons in
  ${p}\overline{p}$ collisions at $\sqrt{s} = 1.96$ TeV}'',} \textit{ Phys.
  Rev. D} \textbf{ 73} (2006) 111102,
  \href{http://dx.doi.org/10.1103/PhysRevD.73.111102}{\doi{10.1103/PhysRevD.73.111102}},
  \href{http://www.arXiv.org/abs/hep-ex/0604040}{\texttt{
  arXiv:hep-ex/0604040}}.

\bibitem{D0estar}
\hrefCMSnoop {} {{ D0} Collaboration, ``{Search for excited electrons in
  ${p}\overline{p}$ collisions at $\sqrt{s} = 1.96$ TeV}'',} \textit{ Phys.
  Rev. D} \textbf{ 77} (2008) 091102,
  \href{http://dx.doi.org/10.1103/PhysRevD.77.091102}{\doi{10.1103/PhysRevD.77.091102}},
  \href{http://www.arXiv.org/abs/0801.0877}{\texttt{ arXiv:0801.0877}}.

\bibitem{LHC}
\hrefCMSnoop {} {L.~Evans and P.~Bryant, ``{LHC Machine}'',} \textit{ JINST}
  \textbf{ 3} (2008) S08001,
  \href{http://dx.doi.org/10.1088/1748-0221/3/08/S08001}{\doi{10.1088/1748-0221/3/08/S08001}}.

\bibitem{cms-limit}
\hrefCMSnoop {} {{ CMS} Collaboration, ``{A search for excited leptons in pp
  collisions at $\sqrt{s}$ = 7 TeV}'',} \textit{ Phys. Lett. B} \textbf{ 704}
  (2011) 143,
  \href{http://dx.doi.org/10.1016/j.physletb.2011.09.021}{\doi{10.1016/j.physletb.2011.09.021}},
\href{http://www.arXiv.org/abs/1107.1773}{\texttt{ arXiv:1107.1773}}.
%%CITATION = ARXIV:1107.1773;%%.

\bibitem{atlas-limit}
\hrefCMSnoop {} {{ ATLAS} Collaboration, ``{Search for excited leptons in
  proton-proton collisions at $\sqrt{s}$ = 7 TeV with the ATLAS detector}'',}
  \textit{ Phys. Rev. D} \textbf{ 85} (2012) 072003,
  \href{http://dx.doi.org/10.1103/PhysRevD.85.072003}{\doi{10.1103/PhysRevD.85.072003}},
\href{http://www.arXiv.org/abs/1201.3293}{\texttt{ arXiv:1201.3293}}.
%%CITATION = ARXIV:1201.3293;%%.

\bibitem{JINST}
\hrefCMSnoop {} {{ CMS} Collaboration, ``The {CMS} experiment at the {CERN}
  {LHC}'',} \textit{ JINST} \textbf{ 03} (2008) S08004,
\href{http://dx.doi.org/10.1088/1748-0221/3/08/S08004}{\doi{10.1088/1748-0221/3/08/S08004}}.
%%CITATION = JINST,3,S08004;%%.

\bibitem{Sjostrand:2006za}
\hrefCMSnoop {} {T.~Sj{\"o}strand, S.~Mrenna, and P.~Z. Skands, ``{PYTHIA} 6.4
  physics and manual'',} \textit{ JHEP} \textbf{ 05} (2006) 026,
  \href{http://dx.doi.org/10.1088/1126-6708/2006/05/026}{\doi{10.1088/1126-6708/2006/05/026}},
  \href{http://www.arXiv.org/abs/hep-ph/0603175}{\texttt{
  arXiv:hep-ph/0603175}}.

\bibitem{Sjostrand:pythia8}
\hrefCMSnoop {} {T.~Sj{\"o}strand, S.~Mrenna, and P.~Z. Skands, ``A brief
  introduction to {PYTHIA 8.1}'',} \textit{ Comput. Phys. Commun.} \textbf{
  178} (2008) 852,
  \href{http://dx.doi.org/10.1016/j.cpc.2008.01.036}{\doi{10.1016/j.cpc.2008.01.036}},
  \href{http://www.arXiv.org/abs/hep-ph/07103820}{\texttt{
  arXiv:hep-ph/07103820}}.

\bibitem{cteq}
J.~Pumplin\hrefCMSnoop {} { {et~al.}, ``{New Generation of Parton Distributions
  with Uncertainties from Global QCD Analysis}'',} \textit{ JHEP} \textbf{ 07}
  (2002) 012,
  \href{http://dx.doi.org/10.1088/1126-6708/2002/07/012}{\doi{10.1088/1126-6708/2002/07/012}},
  \href{http://www.arXiv.org/abs/hep-ph/0201195}{\texttt{
  arXiv:hep-ph/0201195}}.

\bibitem{madgraph}
\hrefCMSnoop {} {J.~Alwall {et~al.}, ``{MadGraph/MadEvent v4: The New Web
  Generation}'',} \textit{ JHEP} \textbf{ 0709} (2007) 028,
  \href{http://dx.doi.org/10.1088/1126-6708/2007/09/028}{\doi{10.1088/1126-6708/2007/09/028}},
\href{http://www.arXiv.org/abs/0706.2334}{\texttt{ arXiv:0706.2334}}.
%%CITATION = ARXIV:0706.2334;%%.

\bibitem{MCFM}
\hrefCMSnoop {} {J.~M. Campbell and R.~K. Ellis, ``{MCFM for the Tevatron and
  the LHC}'',} in \textit{ Loops and Legs in Quantum Field Theory,
  $10^\text{th}$ DESY Workshop on Elementary Particle Theory}.
\newblock 2010.
\newblock \href{http://www.arXiv.org/abs/1007.3492}{\texttt{ arXiv:1007.3492}}.
\newblock Nucl. Phys. Proc. Suppl. 205--206.
\href{http://dx.doi.org/10.1016/j.nuclphysbps.2010.08.011}{\doi{10.1016/j.nuclphysbps.2010.08.011}}.
%%CITATION = ARXIV:1007.3492;%%.

\bibitem{geant4}
\hrefCMSnoop {} {{ GEANT4} Collaboration, ``{GEANT4}---a simulation toolkit'',}
  \textit{ Nucl. Instrum. Meth. A} \textbf{ 506} (2003) 250,
  \href{http://dx.doi.org/10.1016/S0168-9002(03)01368-8}{\doi{10.1016/S0168-9002(03)01368-8}}.

\bibitem{PhoRec}
\hrefCMSnoop {} {{ CMS} Collaboration, ``{Photon Reconstruction and
  Identification at $\sqrt{s}$ = 7 TeV}'',} \textit{ {CMS Physics Analysis
  Summary}} \textbf{ {EGM-10-005}} (2010).

\bibitem{clopperpearson}
\hrefCMSnoop {} {C.~J. Clopper and E.~S. Pearson, ``The Use of Confidence or
  Fiducial Limits Illustrated in the Case of the Binomial'',} \textit{
  Biometrika} \textbf{ 26} (1934) 404,
  \href{http://dx.doi.org/10.1093/biomet/26.4.404}{\doi{10.1093/biomet/26.4.404}}.

\bibitem{zprimetoee}
\href {http://cdsweb.cern.ch/record/1446209} {{ CMS} Collaboration, ``Search
  for Contact Interactions in $\mu^+\mu^-$ Events in pp Collisions at
  $\sqrt{s}$ = 7 {TeV}'',} CMS Physics Analysis Summary CMS-PAS-EXO-11-009,
  2011.

\bibitem{lumi}
\href {http://cdsweb.cern.ch/record/1434360} {{ CMS} Collaboration, ``Absolute
  Calibration of the Luminosity Measurement at {CMS}: {W}inter 2012 Update'',}
  CMS Physics Analysis Summary CMS-PAS-SMP-12-008, 2012.

\bibitem{higgstool}
\href {http://cdsweb.cern.ch/record/1379837} {{ATLAS and CMS Collaborations},
  ``Procedure for the {LHC} Higgs boson search combination in {S}ummer 2011'',}
  Technical Report CMS-NOTE-2011-005, ATL-PHYS-PUB-2011-011, CERN, Geneva,
  2011.

\bibitem{Bayes}
J.~Heinrich\hrefCMSnoop {} { {et~al.}, ``interval estimation in the presence of
  nuisance parameters. 1. {Bayesian} approach'',} (2004).
  \href{http://www.arXiv.org/abs/physics/0409129}{\texttt{
  arXiv:physics/0409129}}.

\bibitem{PDGBayes}
\hrefCMSnoop {} {K.~Nakamura {et~al.}, ``The Review of Particle Physics'',}
  \textit{ J. Phys. G} \textbf{ 37} (2010) 075021,
  \href{http://dx.doi.org/10.1088/0954-3899/37/7A/075021}{\doi{10.1088/0954-3899/37/7A/075021}}.
  Sec. 33.3.1.

\bibitem{CLs}
\hrefCMSnoop {} {A.~L. Read, ``Presentation of search results: the {$CL_s$}
  technique'',} \textit{ J. Phys. G} \textbf{ 28} (2002), no.~10, 2693,
  \href{http://dx.doi.org/10.1088/0954-3899/28/10/313}{\doi{10.1088/0954-3899/28/10/313}}.

\bibitem{Junk:1999kv}
\hrefCMSnoop {} {T.~Junk, ``Confidence level computation for combining searches
  with small statistics'',} \textit{ Nucl. Instrum. Meth. A} \textbf{ 434}
  (1999) 435,
  \href{http://dx.doi.org/10.1016/S0168-9002(99)00498-2}{\doi{10.1016/S0168-9002(99)00498-2}},
\href{http://www.arXiv.org/abs/hep-ex/9902006}{\texttt{ arXiv:hep-ex/9902006}}.
%%CITATION = HEP-EX/9902006;%%.

\end{thebibliography}\endgroup
\cleardoublepage \appendix\section{The CMS Collaboration \label{app:collab}}\begin{sloppypar}\hyphenpenalty=5000\widowpenalty=500\clubpenalty=5000\textbf{Yerevan Physics Institute,  Yerevan,  Armenia}\\*[0pt]
S.~Chatrchyan, V.~Khachatryan, A.M.~Sirunyan, A.~Tumasyan
\vskip\cmsinstskip
\textbf{Institut f\"{u}r Hochenergiephysik der OeAW,  Wien,  Austria}\\*[0pt]
W.~Adam, E.~Aguilo, T.~Bergauer, M.~Dragicevic, J.~Er\"{o}, C.~Fabjan\cmsAuthorMark{1}, M.~Friedl, R.~Fr\"{u}hwirth\cmsAuthorMark{1}, V.M.~Ghete, J.~Hammer, N.~H\"{o}rmann, J.~Hrubec, M.~Jeitler\cmsAuthorMark{1}, W.~Kiesenhofer, V.~Kn\"{u}nz, M.~Krammer\cmsAuthorMark{1}, I.~Kr\"{a}tschmer, D.~Liko, I.~Mikulec, M.~Pernicka$^{\textrm{\dag}}$, B.~Rahbaran, C.~Rohringer, H.~Rohringer, R.~Sch\"{o}fbeck, J.~Strauss, A.~Taurok, W.~Waltenberger, G.~Walzel, E.~Widl, C.-E.~Wulz\cmsAuthorMark{1}
% ...
\vskip\cmsinstskip
\dag:~Deceased\\
1:~~Also at Vienna University of Technology, Vienna, Austria\\
% ...

\end{sloppypar}
\end{document}